\documentclass[english,nofootinbib,notitlepage,prd,twocolumn]{revtex4-1}
\pdfoutput=1

\usepackage[T1]{fontenc}
\usepackage[utf8]{inputenc}
\usepackage{natbib}
\usepackage{amssymb,amsmath}
\usepackage{babel}
\usepackage[normalem]{ulem}

\def\be{\begin{equation}}
\def\ee{\end{equation}}

\makeatletter

\usepackage[pdftex]{graphicx}
\usepackage[pdftex]{color}
\usepackage[colorlinks,bookmarks]{hyperref}
\definecolor{linkblue}{rgb}{0,0,0.8}
\definecolor{linkgreen}{rgb}{0,0.5,0}
\definecolor{darkgreen}{rgb}{0,0.4,0}
\definecolor{purple}{rgb}{0.7,0.0,0.4}
\hypersetup{linkcolor=linkblue, citecolor=purple, urlcolor=linkblue}

\makeatother

\begin{document}

\title{CMB all-scale blackbody distortions induced by linearizing temperature}

\author{Alessio Notari$^{1}$}
\author{Miguel Quartin$^{2}$}

\affiliation{$^{1}$   Departament de F\'isica Fondamental i Institut de Ci\`encies del Cosmos, Universitat de Barcelona, Mart\'i i Franqu\`es 1, 08028 Barcelona, Spain}
\affiliation{$^{2}$ Instituto de Física, Universidade Federal do Rio de Janeiro, 21941-972, Rio de Janeiro, Brazil}

\begin{abstract}
    Cosmic Microwave Background (CMB) experiments, such as WMAP and Planck,  measure intensity anisotropies and build maps using a \emph{linearized} formula for relating them to the temperature blackbody fluctuations.
    However, this procedure also generates a signal in the maps in the form of y-type distortions which is degenerate with the thermal Sunyaev Zel’dovich (tSZ) effect. These are small effects that arise at second-order in the temperature fluctuations not from primordial physics but from such a limitation of the map-making procedure. They constitute a contaminant for  measurements of: our peculiar velocity, the tSZ and primordial $y$-distortions. They can nevertheless be well-modeled and accounted for.  We show that the distortions arise from a leakage of the CMB dipole into the $y$-channel which couples to all multipoles, mostly affecting the range $\ell \lesssim 400$. This should be visible in Planck's $y$-maps with an estimated signal-to-noise ratio of about 12. We note however that such frequency-dependent terms carry no new information on the nature of the CMB dipole. This implies that the real significance of Planck's Doppler coupling measurements  is actually lower than reported by the collaboration. Finally, we quantify the level of contamination in tSZ and primordial $y$-type distortions and show that it is above the sensitivity of proposed next generation CMB experiments.
\end{abstract}


\maketitle

\section{Introduction}

Both WMAP~\cite{Bennett:2012zja} and Planck~\cite{Adam:2015rua} Cosmic Microwave Background (CMB) experiments measured photon intensity anisotropy maps at different frequencies, which were then combined to extract a pure blackbody spectrum, filtering out other signals with different spectra.
However such a procedure has been carried out only at linearized level in temperature fluctuations $\delta T/T$, and in the present work we show that at second-order a $y$-type distortion  in the CMB is  generated, not due to any primordial process, but due to this map-making procedure (for a review on $y$-type distortions see e.g.~\cite{Andre:2013nfa}).  These distortions can (and should) nevertheless be modelled and accounted for in order to remove contaminations from the measured $y$-maps.

Even assuming the CMB to be a pure blackbody in its rest frame, such fake spectral distortions are ``generated'' dominantly by the CMB dipole, since it is by far the largest  CMB temperature fluctuation.
Actually,  \emph{all} first-order perturbation quantities in temperature generate $y$-distortions in the usual map-making procedure at quadratic level. The largest such distortion $(\sim 10^{-6})$ comes of course from the dipole terms squared, which produce quadrupole distortions in the $y$-maps, first discussed in~\cite{Kamionkowski:2002nd}, and monopole distortions~\cite{Chluba:2004cn}. The second largest $(\sim 10^{-8})$, discussed in~\cite{Aghanim:2013suk}, consists in couplings between different multipoles, that arise from cross-terms containing the dipole and the other multipoles. The main goal of this paper is to quantify such $y$-type couplings.

The dimensionless amplitude $\Delta_1$ of the CMB dipole was measured by Planck to be $(1.2345 \pm 0.0007) \times 10^{-3}$~\cite{Adam:2015vua}. This value is understood to be mostly due to the velocity of the observer, \emph{i.e.}~our peculiar velocity.

We note however a fraction of the dipole should be also generated by the dipolar part of the large scale gravitational potential~\cite{Roldan:2016ayx}, at least to ${\cal O}(1\%)$ in a standard scenario. Neglecting this and assuming our velocity to be the only contribution to the CMB dipole, we get $\Delta_1 = \beta\equiv v/c$. A boost has two effects on an image of the sky: Doppler and aberration. While aberration only changes the arrival direction of photons, Doppler affects the frequency spectrum in a direction-dependent way. The Doppler effect is non-trivial even if a map is completely homogeneous in the rest frame, inducing an order $\beta^\ell$ effect on a multipole $\ell$. Since $\beta \sim 10^{-3}$ in practice this affects significantly only the dipole, the quadrupole and the monopole; for the dipole it is the dominant component, for the quadrupole it is a small but non-negligible correction~\cite{Notari:2015kla, Quartin:2015kaa}. Even though the dipole is still a blackbody, it was pointed out originally in~\cite{Kamionkowski:2002nd} that the quadrupole has instead a $y$-type spectrum, and it has been shown that the $y$-type nature of the kinematic quadrupole alters and actually increases the significance of anomalous quadrupole-octupole alignments~\cite{Notari:2015kla} and it could also affect the high frequency calibration of the Planck experiment~\cite{Quartin:2015kaa} (although see~\cite{Aghanim:2016yuo}).

The Doppler effect also induces a coupling between different multipoles in non-homogenous maps. For this purpose we can decompose the CMB primordial temperature in the rest frame as a monopole plus perturbations, dependent on the $\boldsymbol{\hat{n}}$ direction: $T=T_0  + \varepsilon \, \delta T(\boldsymbol{\hat{n}})$, where  we  define $\varepsilon \equiv 10^{-5}$ and so  $ \delta T(\boldsymbol{\hat{n}})$ for large scales is of order unity. On such maps Doppler induces a  $\ell$, $\ell\pm 1$ correlation of order $\beta \varepsilon\sim 10^{-8}$. Aberration induces a $\beta \varepsilon$ coupling between $\ell$ with $\ell\pm n$, which is not a simple function in harmonic space~\cite{Notari:2011sb}, but  the main effect of which in practice is  also a $\ell$, $\ell\pm 1$ correlation~\cite{Challinor:2002zh,Kosowsky:2010jm,Amendola:2010ty,Chluba:2011zh}, which was measured (together with Doppler) by Planck at 2.8$\sigma$~\cite{Aghanim:2013suk}. As we will now show there is  an additional $\ell, \ell\pm1$ correlation created by the map-making procedure which creates $y$-type distortion of the blackbody spectrum, and which also shows up as a Doppler-like $\ell$, $\ell\pm 1$ correlation. Such $y$-distortions were computed also in the context of moving clusters by~\cite{Sazonov:1998ae}; for the quadrupole by~\cite{Kamionkowski:2002nd,Chluba:2004cn,Sunyaev:2013coa}; for the superposition of two blackbodies~\cite{Chluba:2004cn}; as an enhancement of pre-existing $y$-type distortions by~\cite{Balashev:2015lla}. See also~\cite{Challinor:2002zh,  Aghanim:2013suk}.

\section{Side effects of linearizing temperature}

In a given frame (which could be the CMB rest frame or another boosted frame) an observer measures blackbody photons with observed frequency $\nu$ with a specific intensity (or spectral radiance):
\begin{equation}
    I(\nu) = \frac{h}{c^2}\frac{2 \nu^3}{e^{\frac{h \nu }{k_B T(\boldsymbol{\hat{n}})}}-1} \,. \label{eq:Int}
\end{equation}
Here we decompose $T(\boldsymbol{\hat{n}})=T_0+\Delta T(\boldsymbol{\hat{n}})$.
Following CMB conventions, in what follows for simplicity we will refer to specific intensity as just ``intensity'', although technically this latter term usually refers to the bolometric specific intensity. Taylor expanding to first order we get
\begin{equation}
    \delta I(\nu,\boldsymbol{\hat{n}})\,\approx\, \frac{h}{c^2}\frac{2 \nu ^4   e^{\frac{\nu }{\nu_0}}}{T_0^2 \left(e^{\frac{\nu }{\nu_0}}-1\right)^2}   \, \delta T(\boldsymbol{\hat{n}})
    \,\equiv\, K  \, \frac{\Delta T(\boldsymbol{\hat{n}})}{T_0} \,,
    \label{eq:lin-temp-1st-order}
\end{equation}
with $\nu_0 \equiv k_B T_0 / h = (56.79 \pm 0.01) \, {\rm GHz}$~\cite{Fixsen:2009ug}. This approximate equation is commonly used by the CMB collaborations to \emph{define} temperature as $\delta I(\nu,\boldsymbol{\hat{n}})/K(\nu)$, which although not dependent on frequency differs from the real thermodynamic $T$. Following~\cite{Notari:2015kla} we refer to $L(\boldsymbol{\hat{n}})\equiv \delta I(\nu,\boldsymbol{\hat{n}})/K$ as the \emph{linearized temperature}.

We stress however that one should not stop the above expansion at first order, because second-order terms are non-negligible. Extending~\eqref{eq:lin-temp-1st-order} to second order, we get
\begin{align}\label{eq:lin-temp-2nd-order}
    L(\nu,\boldsymbol{\hat{n}}) \,=\, \frac{\Delta T(\boldsymbol{\hat{n}})}{T_0} + \left( \frac{\Delta T(\boldsymbol{\hat{n}})}{T_0} \right)^2 Q(\nu)\,,
\end{align}
where
\begin{equation}
  Q(\nu) \,\equiv\, \frac{\nu}{2\nu_0}  \coth\left[\frac{\nu}{2\nu_0}\right] .
\end{equation}
The second order term in~\eqref{eq:lin-temp-2nd-order} tells us that \emph{any} first order perturbation would appear as second-order blackbody distortions in the CMB~\cite{Aghanim:2015eva}. In particular, this specific frequency dependency is called a $y$-type distortion and is degenerate with the thermal Sunyaev Zel'dovich effect (tSZ) effect~\cite{Kamionkowski:2002nd}. In what follows  we quantify such effects for Planck and future experiments. Of course such effect could be removed simply by solving eq.~\eqref{eq:lin-temp-2nd-order} for the variable $\,\Delta T(\boldsymbol{\hat{n}})/T_0$. However, since this has not been done in the WMAP or Planck map-making procedure, one should be aware that when analyzing $y$-type maps, part of the signal is contaminated by this $Q(\nu)$-dependent term. In the rest of this manuscript we quantify such effects for Planck and future experiments.

In an arbitrary reference frame the Doppler term of order $\beta$ contributes to the CMB dipole, whose amplitude $\Delta_1$ is a sum of two terms:  $\Delta_1\sim \varepsilon + \beta$, which can be much larger than $\varepsilon$ (on the Sun's frame we have $\Delta_1\sim 10^{-3}$). Using $\,\mu=\boldsymbol{\hat{\Delta}_1}\cdot \boldsymbol{\hat{n}}$, where $\,\boldsymbol{\hat{\Delta}_1}\,$ is the direction of the dipole, we can split $\,\Delta T/T\,=\,\Delta_1 \mu + \delta T/T\,$ and rewrite \eqref{eq:lin-temp-2nd-order} as:
\begin{align}\label{eq:L-2nd-order-full}
    L(\nu,  &\,\boldsymbol{\hat{n}})  \,=\, \mu \Delta_1  +\varepsilon \frac{\delta T}{T_0} - \frac{1}{2} \tilde{\beta}^2 - \mu \varepsilon \tilde{\beta} \, \frac{\delta T}{T_0} + \varepsilon \tilde{\beta} \left(\frac{\delta T_{ab}}{T_0}    \right)   \nonumber \\
    &+ \!\Bigg[\!\left( \mu^2-\frac{1}{3}\right)\! \Delta_1^2 +\frac{1}{3} \Delta_1^2  + 2\varepsilon  \Delta_1  \mu \frac{\delta T}{T_0} \Bigg] Q(\nu)   \nonumber \\
    & + L_{\rm higher}
    \, .
\end{align}
Above $\delta T$ refers to first-order temperature anisotropies for $\ell\ge2$, $(\delta T_{ab} / T_0)$ refers to the aberration terms and $\tilde{\beta}$ refers to the contributions due to our peculiar velocity (although this quantity in reality contains also some terms due to intrinsic cosmological perturbations, as discussed below). We have kept only leading order terms of order $\varepsilon \Delta_1 \cdot \delta T/T$, so $L_{\rm higher}$ stands for terms of order $\Delta_1^3$ or higher (i.e., including terms of order $\varepsilon^2$). This expansion is in agreement with~\cite{Aghanim:2013suk}.

Note that all second order terms which are \emph{not} proportional to $Q(\nu)$ are in fact true temperature fluctuations due to a boost, contained in the first term of eq.~\eqref{eq:lin-temp-2nd-order}. In particular in the second line in the above equation, the first term is the frequency-dependent Doppler-quadrupole (DQ) discussed in~\cite{Kamionkowski:2002nd} and the second is a $y$-type monopole, analyzed in~\cite{Chluba:2004cn}.  In the original version of~\cite{Kamionkowski:2002nd} it was hoped the DQ could be used to measure our velocity, but the authors later understood it could not disentangle the Doppler contributions of order $\beta$ from the intrinsic dipole of order $\varepsilon$. From~\eqref{eq:lin-temp-2nd-order} and~\eqref{eq:L-2nd-order-full} this is clear: no matter what is behind $\Delta_1$ the $Q(\nu)$ distortions are the same. The last term in the second line is also generated by the map-making procedure, so it carries no new information about $\beta$.

\subsection{Terms affected by our peculiar velocity}\label{sec:beta}

The terms proportional to $\tilde{\beta}$ in eq.~\eqref{eq:L-2nd-order-full} are the ones physically generated directly by a Lorentz boost due to our peculiar velocity, and not from a leakage of the total dipole. We nevertheless use $\tilde{\beta}$ instead of $\beta$ because it can contain also a contribution due to second order effects of an intrinsic large-scale mode of the gravitational potential. In fact, as discussed in more detail in~\cite{Roldan:2016ayx} such a mode produces both aberration and Doppler couplings. While the aberration couplings can only be mimicked by a fine-tuned gravitational potential, Doppler couplings are naturally produced in a way which is exactly degenerate with a boost in the case of Gaussian initial conditions from inflation, even in the absence of a peculiar velocity. Nevertheless, other primordial scenarios do not produce such couplings, so measuring the Doppler couplings and comparing them to the dipole and to aberration can tell us both about the primordial universe and about our peculiar velocity.
The 4$^{\rm th}$ term in eq.~\eqref{eq:L-2nd-order-full} therefore contains the physical effect of a genuine Doppler coupling due to our velocity and can be used to measure $\beta$.

However, one should \emph{not} construct an estimator aimed at
measuring the sum of the 4$^{\rm th}$ and last terms of eq.~\eqref{eq:L-2nd-order-full} because the latter does not necessarily come from a boost and so it may tell us nothing about the physical nature of the dipole. Unfortunately, this is precisely what was done in~\cite{Aghanim:2013suk}, where the $Q(\nu)$ terms were also considered under the name of \emph{boost factors}. The collaboration was nevertheless aware that part of the signal in their estimators would have ``arisen in the presence of any sufficiently large temperature fluctuation''~\cite{Aghanim:2013suk}, but they did not conduct a separate analysis removing the $Q(\nu)$ terms. As a consequence their measurement has in reality slightly less significance than the value that was quoted, because the estimator should not have been multiplied by the boost factors if one wants a truly physically independent measurement of our velocity. Instead the optimal procedure is to remove the $Q(\nu)$-dependent terms and then measure the couplings.

Measuring the $Q(\nu)$ terms can serve only the purpose of a cross-check, as we discuss below in Section~\ref{sec:detecting-DD}. Moreover in the case in which the analysis is carried directly on component-separated CMB maps, the $Q(\nu)$ could have already have been projected out, in which case no average boost factor should be included. This is not the case for the analysis in~\cite{Aghanim:2013suk}  using the CMB maps, because these maps did not project out the tSZ signal, as we show below.

Clearly, the $Q(\nu)$ terms can also appear as contamination on the tSZ measurements and of primordial $y$-distortions, which rely on the same channel. We address all these issues below in Section~\ref{sec:contamination}.

\subsection{The all-scale Dipolar Distortion}

There are two terms proportional to  $Q(\nu)$  in~\eqref{eq:L-2nd-order-full}:
\begin{align}\label{eq:y-DQ}
  2  Q(\nu) \frac{ \mu^2 \Delta_1 ^2}{2} \,\equiv\,  y_{\rm DQ} \,, \qquad
  2  Q(\nu)  \mu \Delta_1 \varepsilon  \, \frac{\delta T}{T_0} \,\equiv \, y_{\rm DD} .
\end{align}
The $y_{\rm DQ} (\boldsymbol{\hat{n}})$ and  $y_{\rm DD} (\boldsymbol{\hat{n}})$ correspond to conventional $y$-type distortion maps. The former was thoroughly discussed in~\cite{Kamionkowski:2002nd}. The latter term we refer to as an all-scale dipolar  (or Doppler) distortion (shorthand ``DD''), since it affects all multipoles~\cite{Chluba:2004cn} and since the dipole is supposed to be mostly due to Doppler. In this paper we stress that these dipolar distortions should be visible in Planck's data, as discussed below. We will elaborate the consequences of this term in what follows; from~\eqref{eq:lin-temp-2nd-order} and~\eqref{eq:L-2nd-order-full}, however, we remind again that the DD are insensitive to the origin of the dipole and thus, just like the Doppler quadrupole, it cannot be used to measure our peculiar velocity independently of the temperature dipole.

From the above equation the DD coefficients of the $y_{\rm DD} (\boldsymbol{\hat{n}}) $ map can be written in multipole space as a function of the $a^{\rm T}_{\ell m}$ (the harmonic coefficients of $\delta T(\boldsymbol{\hat{n}})/T_0$)
as~\cite{Challinor:2002zh,Amendola:2010ty,Quartin:2014yaa}:
\begin{equation}
    a^{\rm DD}_{\ell m}=  \Delta_1 \big(G_{\ell, m} a^{\rm T}_{\ell-1 m} + G_{\ell+1, m} a^{\rm T}_{\ell+1 m}\big)\,, \label{signalDD}
\end{equation}
with $G_{\ell, m}\equiv \sqrt{(\ell^2-m^2)/(4 \ell^2-1)}$.

As a consequence the $a^{\rm DD}_{\ell m}$ coefficients can be predicted since the $a^{\rm T}_{\ell m}$ are known by the temperature maps. Note also that the above equation assumes  the dipole to be along the $\boldsymbol{\hat{z}}$ axis, when making the harmonic decomposition: we first discuss this simplified framework and then discuss the  general case.

\section{Detecting the DD in the $y$-maps: a consistency check} \label{sec:detecting-DD}

As stressed in the Introduction one could now perform a consistency check, trying to detect the DD signal in the $y$-maps. In other words one could measure $\Delta_1$ on such maps, without using information from the measurement of the usual blackbody dipole. For this purpose one can first measure the $a^{\rm T}_{\ell m}$ (with $\ell>2$)  by building a map which contains  the pure blackbody signal, obtained  combining in a suitable way  the different intensity channels of an experiment (such as Planck) and in this way  we can compute the ${a}^{\rm DD}_{\ell m}$ coefficients using eq.~(\ref{signalDD}). Subsequently we can build a second map of the signal proportional to $Q(\nu)$ by using a different linear combination of the frequency channels and look on this map for such expected ${a}^{\rm DD}_{\ell m}$. In this way we provide a consistency check which we can rephrase as a measurement of $\Delta_1$, which is the only free parameter in ${a}^{\rm DD}_{\ell m}$.
Note that for this purpose we will treat the  tSZ, which has the same $Q(\nu)$ dependence, as a noise. Instead in the next section we will do the opposite: check what is the noise generated by the DD on tSZ maps.

Let us assume that the CMB is made of $N$ different signals, which can include cosmological signals as well as foregrounds and noise, such that the linearized temperature is $L = \sum_n \alpha_n L_n(\nu,\boldsymbol{\hat{n}})$. For instance here the CMB  blackbody signal $L_{\rm CMB}$ would be only a function of $\boldsymbol{\hat{n}}$, flat in $\nu$, coming from the first line of eq.~\eqref{eq:L-2nd-order-full}:
\begin{equation}
    L_{\rm CMB} \,\equiv\,   \, \frac{\delta T(\boldsymbol{\hat{n}})}{T_0}+ \mu\Delta_1    -   \frac{1}{2} \tilde{\beta}^2 - \mu \varepsilon \tilde{\beta}  \, \frac{\delta T(\boldsymbol{\hat{n}})}{T_0}   \, . \label{eq:LCMB}
\end{equation}
Given an experiment which has  different channels $L(\nu_k)$,  $k=1,...,  K$ (for Planck $K=9$) we can combine them with some weights $w_i$ and build arbitrary combinations (maps $M$):
\begin{equation}
    M \,=\, \sum_{{\rm channel } \; i} w_i L(\nu_i) \, .
\end{equation}
To fix the weights we need to specify $K$ constraints. If for instance we want to project out the CMB blackbody signal we can build a map with $\sum_i w_i =0$, while if we wanted to project out the $y$-type signal  we should impose \begin{equation}\label{eq:wQ=0}
    Q_{\rm eff} \,\equiv\, \sum_i w_i Q(\nu_i) \,=\, 0 \,.
\end{equation}
This was not done however in Planck CMB maps, as we discuss in more detail in  Appendix~\ref{app:planck-Qw}.

In general, the procedure above can be used to project out several linearly independent signals, as long as they can be factorized as a frequency-dependent function times an angular-dependent function.  At most we can project out $K-1$ signals (one constraint is the overall normalization of the map which has to be  fixed). As already mentioned the tSZ is also proportional to $Q(\nu)$:
\begin{equation}
    L_{\rm tSZ}(\nu,\boldsymbol{\hat{n}})=\big(2Q(\nu)-4\big) y_{\rm tSZ} (\boldsymbol{\hat{n}})\,,
\end{equation}
so that the DD is a linear combination of a pure CMB signal and a pure tSZ signal.

We consider then a $y$-projected map $M^{y}(\boldsymbol{\hat{n}})$, in which the CMB and other foregrounds are projected out. Such maps have been already constructed for Planck~\cite{VanWaerbeke:2013cfa, Hill:2013dxa,Aghanim:2015eva}. Their harmonic coefficients are then a sum of three terms
\begin{equation}
    a^y_{\ell m} \,=\, a^{\rm tSZ}_{\ell m} + a^{\rm DD}_{\ell m} + n^y_{\ell m} \, ,
\end{equation}
where $n^y_{\ell m}$ is a noise signal on such a map, with spectrum $N^{yy}_\ell\equiv \langle |n^y_{\ell m}|^2 \rangle$.  Note that at the level of the angular power spectrum $C^{yy}_{\ell}\equiv \sum_m |a^y_{\ell m}|^2 / (2\ell+1)$ the DD  gives only a tiny $\Delta_1^2$ correction in full-sky maps, similar to what happens in the CMB maps~\cite{Catena:2012hq}, but it is clearly visible at the level of the individual $a_{\ell m}$'s as follows.\footnote{Alternatively one could look at 2-point functions of the form $\langle a^y_{\ell \, m} a^y_{\ell+1 \, m}\rangle$ as in~\cite{Kosowsky:2010jm,Amendola:2010ty,Notari:2011sb}.}
As we said, since we are focusing on a detection of the DD signal in the $y$-maps we treat here the tSZ signal as  noise. Following~\cite{Amendola:2010ty} we define
\begin{equation}
    a^{\rm DD}_{\ell m} \;\equiv\; \Delta_1 \hat{a}^{\rm DD}_{\ell m}
\end{equation}
and build a $\chi^2$ from which we compute the signal-to-noise ratio. Since the DD affect both real and imaginary parts of the temperature $a_{\ell m}$'s, we can treat both these terms independently, and write:
\begin{align}
   \chi^2\, \;=\;& \,\sum_{\ell=3}^{\ell_{\rm max}} \sum_{m=0}^{\ell} \frac{{\rm Re}[\bar{a}^{y}_{\ell m} - \Delta_1 \hat{a}^{\rm DD}_{\ell m} ]^2}{\sigma^2_\ell \,(1+\delta_{m 0})/2} \, \nonumber \\
   & + \,\sum_{\ell=3}^{\ell_{\rm max}} \sum_{m=1}^{\ell} \frac{{\rm Im}[\bar{a}^{y}_{\ell m} - \Delta_1 \hat{a}^{\rm DD}_{\ell m} ]^2}{\sigma^2_\ell/2}\,,  \label{chisq}
\end{align}
where $\bar{a}^{y}_{\ell m} $ are the measured harmonic coefficients of the $y$-projected map and $\sigma_\ell^2=N^{yy}_\ell+C^{\rm tSZ}_{\ell}$ in the case in which we consider, for simplicity, the tSZ to be Gaussian.
From this, we can estimate the signal-to-noise ratio directly as:
\begin{align}
   \left(\frac{S}{N}\right)^2 &=\, \sum_{\ell=3}^{\ell_{\rm max}} \sum_{m=0}^{\ell} \frac{{\rm Re}[a^{\rm DD}_{\ell m}]^2}{\sigma^2_\ell (1+\delta_{m 0})/2}+\sum_{\ell=3}^{\ell_{\rm max}} \sum_{m=1}^{\ell} \frac{{\rm Im}[a^{\rm DD}_{\ell m}]^2}{\sigma^2_\ell/2} \nonumber \\
   &=\, \sum_{\ell=3}^{\ell_{\rm max}} \sum_{m=0}^{\ell} \frac{|a^{\rm DD}_{\ell m}|^2}{\sigma^2_\ell} (2-\delta_{m 0})\, .
   \label{SNideal}
\end{align}
The relative estimator can be built by minimizing the $\chi^2$, which leads to:
\begin{equation}
    \hat{\Delta}_1 \;=\;
    \frac{\sum_{\ell =3}^{\ell_{\rm max}} \sum_{m=0}^{\ell}
    {\rm Re} [\bar{a}^{y}_{\ell m}  \cdot \hat{a}_{\ell m}^{\rm DD \, \ast}]/(\sigma^2_\ell (1+\delta_{m 0})) }
    {\sum_{\ell =3}^{\ell_{\rm max}} \sum_{m=0}^\ell
    \big| \hat{a}^{\rm DD}_{\ell m} \big|^2/(\sigma^2_\ell (1+\delta_{m 0})) } \,. \label{estimator}
\end{equation}
Note also that we have omitted the quadrupole from this sum, which would have an ${\cal O}(\tilde{\beta} \epsilon)$ coefficient that is not predictable, since $a_{10}^{\rm T}$ is not known, being dominated by the velocity itself. Moreover we have already seen that  the quadrupole has an  ${\cal O}(\Delta_1^2)$ term in eq.~\eqref{eq:L-2nd-order-full}, which we also discuss separately below.

Equation~\eqref{SNideal} gives the exact DD signal-to-noise ratio, and is the one we used to compute our results. It is nevertheless useful to consider the following simple approximation for the total DD signal. The overall signal contained in each $\ell$ (after summing over $m$) can be obtained by first noting, following~\cite{Notari:2011sb}, that the average value over $m$'s of the $G_{\ell, m}$ coefficients is roughly $0.39$ and its root mean square is roughly~$0.41$. Now, $\langle |a^{\rm T}_{\ell-1 m} + a^{\rm T}_{\ell+1 m}|^2\rangle \simeq C^{\rm TT}_{\ell-1} + C^{\rm TT}_{\ell+1} \simeq 2 C^{\rm TT}_\ell$. Note also that there are $\ell+1$ non-negative $m$'s for each $\ell$, but the $m\neq0$ terms count as double due to the $(2-\delta_{m0})$ term in~\eqref{SNideal}. Substituting these approximations into~\eqref{signalDD} we arrive at the following the average DD signal:
\begin{equation}
   S_{\rm DD}(\ell) \,\equiv\, 0.41 \,\Delta_1\, \sqrt{2\ell+1}\sqrt{2 C^{\rm TT}_{\ell}} \,, \label{DDsignalapprox}
\end{equation}
and thus
\begin{align}
   \left(\frac{S}{N}\right)^2 &\;\simeq\;\, \sum_{\ell=3}^{\ell_{\rm max}} \left[\frac{S_{\rm DD}(\ell)}{\sigma_\ell}\right]^2\, .
   \label{SNapprox}
\end{align}
This approximation yields very similar results to the full calculation, and allows for a better understanding on the dependence of the DD on the different multipoles.

\begin{figure}[t]
    \centering
    \includegraphics[width=0.496 \textwidth]{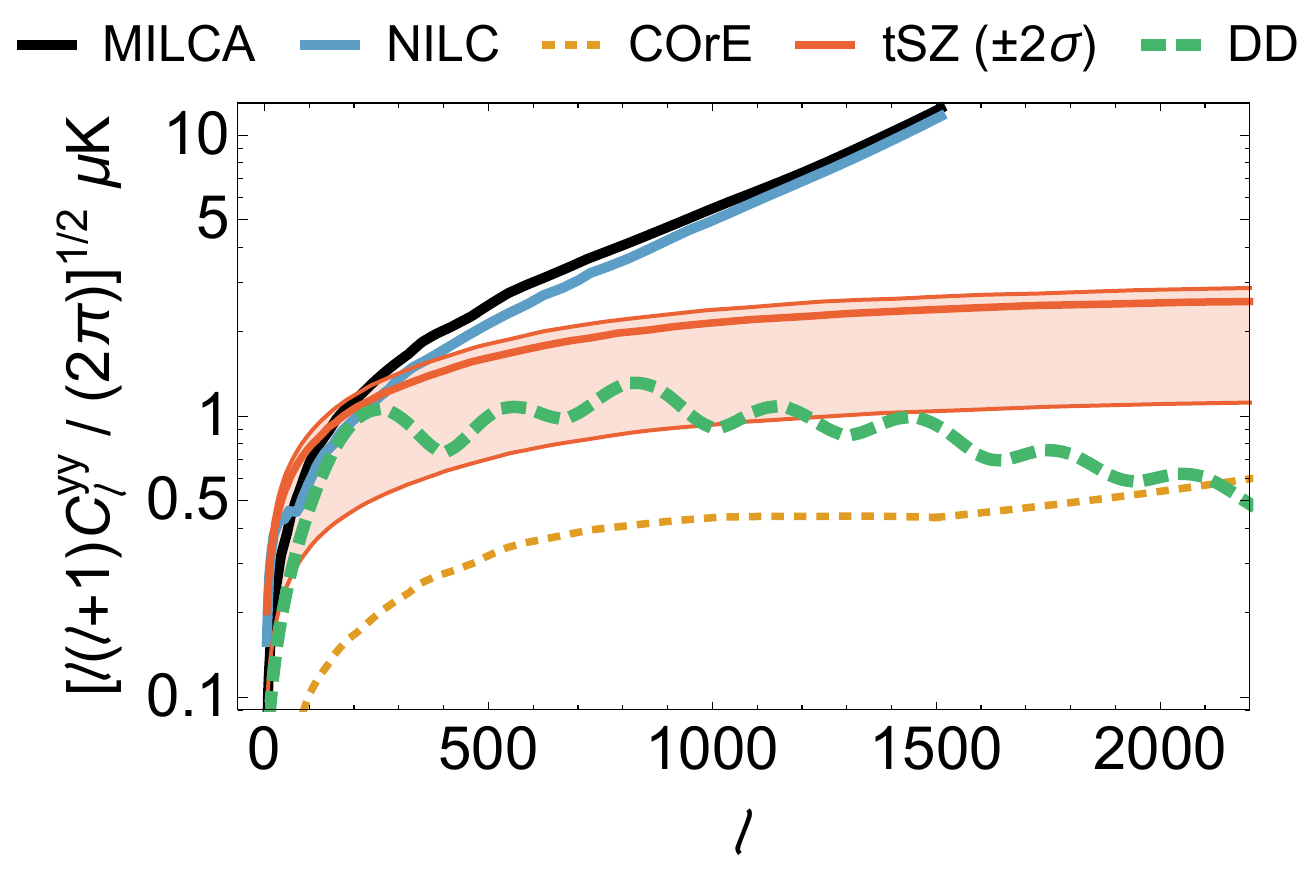}
    \caption{Comparison of spectra between tSZ \emph{signal} and Planck MILCA and NILC \emph{noise}. For the tSZ signal, we show both the best-fit and the $2\sigma$ region allowed by present data~\cite{Aghanim:2015eva}. In dashed green we plot the approximate overall Doppler distortion signal given by $S_{\rm DD} \sqrt{\ell (\ell+1)/(2\pi)}$ [see~eq.~\eqref{DDsignalapprox}]. In dotted yellow we show the noise of the proposed COrE satellite, which lies well below the tSZ signal. \label{fig:tSZ}}
\end{figure}

An estimate for the tSZ spectrum can be taken from Fig.~2 of~\cite{Hill:2013dxa} or from Fig.~17 of~\cite{Aghanim:2015eva}. An estimate for the $N^{yy}_\ell$ noise for Planck can be taken from Fig.~5 of~\cite{Aghanim:2015eva}, which relies on the tSZ-projected maps constructed using either. In Figure~\ref{fig:tSZ} we combine these estimates, together with an estimate of the total DD signal in each $\ell$ as per eq.~\eqref{DDsignalapprox}. Note that current experimental noise is still of the same order of the best-fit tSZ templates, but future experiments such as COrE~\cite{Bouchet:2011ck} will have noise levels well below them (we assume here a resolution of 4 arcmin and a conservative $2\,\mu$K arcmin noise level).

\begin{figure}[t]
    \centering
    \includegraphics[width=0.485\textwidth]{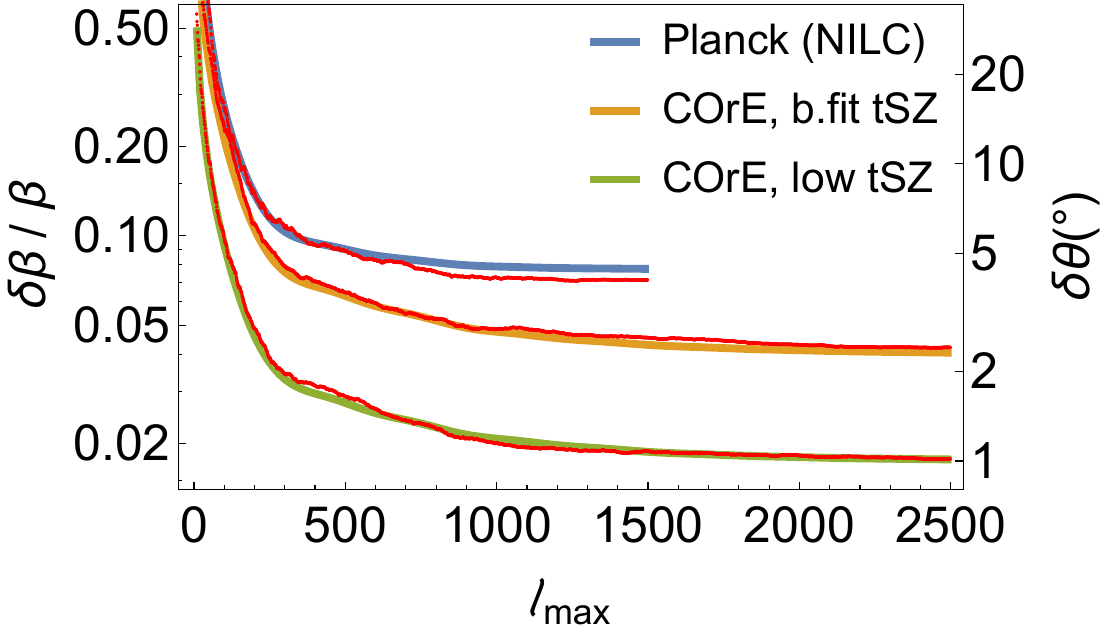}
    \caption{Achievable precision for measuring the dipole amplitude $\Delta_1$ with both Planck (top curve) and the proposed COrE satellite (middle and bottom curves). For Planck we assume the noise levels obtained with the MILCA component separation method.  For COrE we depict the signal corresponding to both the best fit and the $2\sigma$ lowest value of $C_\ell^{\rm tSZ}$ (see Figure~\ref{fig:tSZ}). The thin red curves represent the average of 300 simulations used as cross-check to eq.~\eqref{SNideal}. On the right side we show the precision on the direction, using that (see~\cite{Amendola:2010ty}) $\delta\Delta_1/\Delta_1=\delta\theta$ (in radians). \label{fig:SN-SZ}}
\end{figure}

In Figure \ref{fig:SN-SZ} we depict the achievable precision $\delta\Delta_1$ in the inferred value of the dipole by both Planck and COrE  (or  COrE+, PRISM~\cite{Andre:2013nfa}, PIXIE~\cite{Kogut:2011xw} or any full-sky experiment with considerably  smaller noise than the tSZ spectrum: $N^{yy}_\ell \ll C_\ell^{\rm tSZ}$), using such a consistency check.  Since in this case the detection is only limited by $C_\ell^{\rm tSZ}$, it depends directly on its amplitude. We thus consider two cases: the best fit and the $2\sigma$ lower bound amplitudes of the tSZ template as given by Fig.~17 of~\cite{Aghanim:2015eva}.

We also have built ideal full-sky simulations of the TT and of the $y$-maps, and we have added the DD effect to the latter maps using eq.~(\ref{signalDD}), with a value of $\Delta_1$ as given by the measured dipole. We have run the estimator in eq.~(\ref{estimator}) on 300 simulations (for each case) and we plot in Figure~\ref{fig:SN-SZ} the standard deviation over the mean of the reconstructed  value for $\Delta_1$, as a cross-check of eq.~\eqref{SNideal}. Even for Planck the significance is estimated to be very high, at around $12\sigma$.
Note that for Planck there is almost no signal after $\ell\gtrsim400$ since the noise starts increasing very rapidly while the signal slightly decreases, as can be seen from Figure \ref{fig:tSZ}; for COrE the situation is similar because, even if the noise is negligible, the ratio between the tSZ contamination and the DD signal increases for $\ell \gtrsim 400$ and thus the signal-to-noise grows slowly with $\ell_{\rm max}$ for $\ell_{\rm max} \gtrsim 400$.

The extension of our estimators to the case of a generic direction of the dipole $\boldsymbol{\Delta_1}\equiv (\Delta_{1 \, x}, \Delta_{1 \, y},\Delta_{1 \, z})$ was derived to first order in eq.~(6.1) of~\cite{Quartin:2014yaa}.
In this case, the $\chi^2$ depends linearly on the 3 components of the dipole and by minimization it is straightforward to obtain the estimators $\Delta_{1 \, x},\Delta_{1 \, y}$ and $\Delta_{1 \, z}$. The absolute uncertainty on each single component is given by the exact same estimate of Figure~\ref{fig:SN-SZ} and this can also be translated on an uncertainty on the direction angle by the simple relation $\delta\Delta_1/\Delta_1=\delta\theta$, as discussed already in~\cite{Amendola:2010ty}.

We have also tested in depth whether the dropped ${\cal O}(\Delta_1^2)$ terms could produce any bias to our estimator by including them in the simulations described above.
We got no discernible bias on the inferred $\Delta_1$ nor any change in the scatter (as illustrated by the red curves in Figure~\ref{fig:SN-SZ}). These terms can thus be safely ignored here. For completeness we give the coefficients of $a^{{\rm DD} \, (2)}_{\ell m}$ derived assuming $\Delta_1=\beta$ and expanding eq.~\eqref{eq:Int} to ${\cal{O}}(\beta^2)$
\begin{eqnarray} \label{2nd-order-coefs}
    \hat{a}^{\rm DD \, (2)}_{\ell m}&=& \beta^2 \big(d^0_{\ell m} a^{\rm T}_{\ell m} + d^-_{\ell m} a^{\rm T}_{\ell-2 m} + d^+_{\ell m} a^{\rm T}_{\ell+2 m}\big)\,,   \nonumber \\
    d^0_{\ell m}&=& P(\nu/\nu_0) \, (G_{\ell,m}^2+G_{\ell+1,m}^2)+\frac{1}{2}-Q(\nu )\nonumber \,, \\
    d^+_{\ell m}&=& P(\nu/\nu_0)\, G_{\ell-1,m} G_{\ell,m}  \nonumber \,,  \\
    d^-_{\ell m}&=& P(\nu/\nu_0)\, G_{\ell+1,m} G_{\ell+2,m} \,,
\end{eqnarray}
with $P(x)\equiv \left[2 x+x \cosh (x)-2 \sinh (x)\right]x e^x/(e^x-1)^2$. Note that such tiny ${\cal O}(\beta^2)$ effects exhibit a different frequency dependence from the tSZ.

\begin{figure*}[t]
    \centering
    \includegraphics[width=1.0\textwidth]{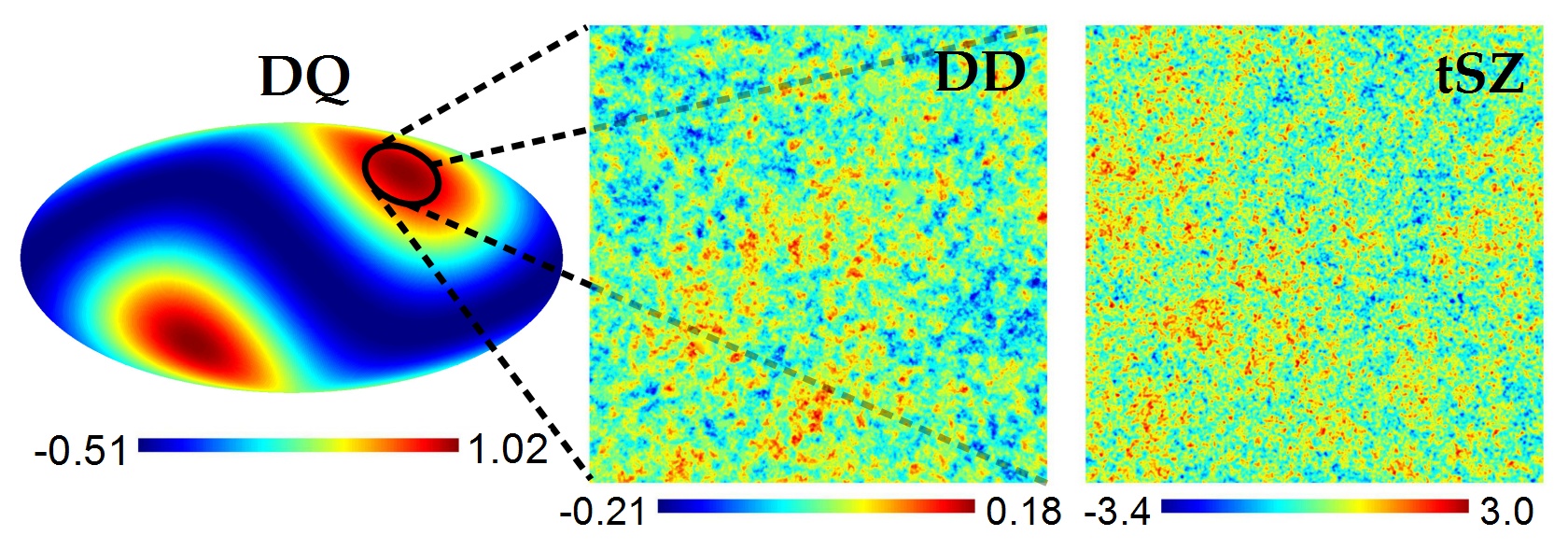}
    \caption{Maps for comparison between the Dipolar Distortions and the tSZ. \emph{Left:} the DQ leakage in the full sky, in galactic coordinates. \emph{Middle:} the DD leakage for a $40^\circ \times 40^\circ$ region (in a Gnomonic projection) around the dipole direction, where the effect is largest. \emph{Right:} same as the middle panel for the simulated tSZ, according to the best-fit Planck $C^{\rm tSZ}_{\ell}$. As can be seen, even around the dipole the DD bias is small: less than 7$\%$ (15$\%$) assuming the current best-fit ($2\sigma$ lower-limit) $C^{\rm tSZ}_{\ell}$. It~can be ignored in Planck data, but not in future experiments like CORE+, PIXIE and PRISM, where it can be higher than the forecast noise. \label{fig:DD-vs-tSZ}}
\end{figure*}

So far we have focused on the ${\cal O}(\varepsilon \Delta_1)$ effects in eq.~\eqref{eq:L-2nd-order-full}. Let us come back again to the DQ term in eq.~\eqref{eq:L-2nd-order-full}, which was shown to be measurable in~\cite{Kamionkowski:2002nd}. In  principle this can seen by Planck: different map-making techniques for extracting the blackbody signal have in fact different combinations of frequencies. Each map $M^{(L)}$ (SMICA, NILC, SEVEM and Commander, in the case of Planck~\cite{Adam:2015tpy}) has weights $w^{L}_i$. Following eq.~\eqref{eq:wQ=0}, this leads to different effective $Q^{L}_{\rm eff}$ terms for the different maps.
So, when subtracting any two of such maps $M^{(1)}$ and $M^{(2)}$ a quadrupole remainder should be observed proportional to $\Delta Q_{\rm eff} \equiv Q^{(1)}_{\rm eff}-Q^{(2)}_{\rm eff}$.

We stress here  that such a remainder also contains a signal which allows us to make a consistency check, giving a measurement of $\Delta_1$ and in the next section we will show how this can contaminate a tSZ map. The Planck Collaboration is aware of this, but they have not conducted such a check explicitly on the grounds that it would require a better understanding of the quadrupole foregrounds~\cite{Aghanim:2013suk}. We can estimate the S/N of the DQ on a temperature map as    $\left( S/N \right)_{\rm DQ} \,=\, 2.725\,Q_{\rm eff} \,(\beta^2/2)/N_{\ell=2}$.

The most straightforward estimation is to use directly a tSZ-projected map, for which $Q_{\rm eff}\equiv 1$ by construction. For Planck we get that, according to Figure~\ref{fig:tSZ}, we can estimate it as $(S/N)_{\rm DQ} \,\simeq\, \rm{17 \;(MILCA) \;\;or\;\; 16 \; (NILC)}\,$.

\section{DD as a contamination to tSZ measurements and primordial $y$-distortions}\label{sec:contamination}

We now discuss how the DD can contaminate the standard tSZ measurements.  The DD effects are maximal close to the dipole direction and its antipode. Figure~\ref{fig:DD-vs-tSZ} depicts both DD and tSZ maps in a region around the dipole direction. The DD is expected to be just a small, $\sim7\%$ effect which is added to the tSZ maps. This is well below Planck's instrumental noise, which is above the $\sim 50 \%$ level at all scales. But it is above the expected CORE+ noise levels, and should therefore be subtracted in the future. At the level of the power spectrum the corrections are tiny, ${\cal O}(\Delta^2_1)$, because they mediate to zero in full-sky. We also show in the full-sky map the amplitude of the DQ in a $y$-map. Let us also note here that such distortions, if not properly accounted for, could in principle affect also measurements of intrinsic spectral distortions. For instance PIXIE~\cite{Kogut:2011xw} should measure with a sensitivity of $10^{-9}$ primordial $y$ distortions, which are expected to be produced at recombination at the $10^{-7}$ level.

When measuring such distortions in the monopole it is certainly necessary to remove the $Q(\nu)$ monopole in eq.~(\ref{eq:L-2nd-order-full}), which is ${\cal O}(10^{-7})$, as noted in~\cite{Chluba:2004cn}. However it is also relevant to remove the DD and DQ distortions. In fact, when introducing a mask, any multipole could leak into the monopole. Moreover one could be also interested in adapting future experiments like PIXIE~\cite{Kogut:2011xw} and PRISM~\cite{Andre:2013nfa} to measure directly spectral distortions of $\ell \geq 1$ multipoles (see~\cite{Balashev:2015lla} for the dipole). In this case a removal of DD and DQ is necessary because it is higher than the instrument sensitivity. Finally let us note that in~\cite{Chluba:2004cn} $y$-distortions due to effects of order $\varepsilon^2\approx 10^{-10}-10^{-11}$ were also studied.

The presence of a mask enhances the leakages into the $y$-channel. This is proportional to the asymmetry of the mask. For the power spectrum there will be an ${\cal O}(\Delta_1)$ leakage on the $C^{yy}_\ell$ (similar to~\cite{Pereira:2010dn}): $\delta C^{yy}_\ell\approx 2\Delta_1\langle \cos\theta\rangle (C_\ell^{\rm TT} C_{\ell}^{\rm tSZ})^{1/2}$, where $\langle \cdot \rangle$ is an angular average over the masked sky. Assuming a mask asymmetry of $10\%$, there would be a $2\%$ contamination at $\ell\le 15$, decreasing at higher $\ell$. For small-sky experiments such as ACT ($\langle \cos\theta\rangle = 0.51$~\cite{Notari:2013iva,Jeong:2013sxy}) the bias is larger, about $2\%$ at $\ell\approx 1000$. For the maps, such a mask would also induce a $10^{-7}$ leak from the DQ and a $10^{-9}$ leak from the DD, which could affect the measurements of the $y$-distortions in the monopole. We stress however that such leakages can be easily avoided by the use of a symmetric mask, as proposed in~\cite{Quartin:2014yaa}, or by subtraction using eq.~\eqref{signalDD}.

\section{Discussion}

We have shown that using a linearized formula for extracting the temperature fluctuations from intensity, one always also induces a leakage on the $y$-maps. Such a signal is dominated by a leakage of the Dipole, the amplitude of which is $\Delta_1\approx 10^{-3}$ and it contains in addition to the known  quadrupole (DQ) and monopole of order $\Delta_1^2$ also a signal proportional to the blackbody temperature map times a dipolar modulation of order $\Delta_1$ over the whole sky, which comes from  a cross-term between the Dipole and the rest of the map. Using the information from the temperature blackbody map, we are able to predict precisely the latter signal (which we called DD) at the level of the individual $a^y_{\ell m}$'s. As we have shown the DD should be present already in Planck at about $12\sigma$ and future experiments are only limited by the degeneracy with the tSZ signal. Detecting this type of signal constitutes a consistency check of the map-making procedure. We also pointed out that the measurement performed in~\cite{Aghanim:2013suk} should have {\it first} removed the $y$-type part of the signal, which is not carrying information independent from the CMB dipole, and then should have measured the blackbody Doppler couplings which are truly induced by a boost. Applying such a procedure will lead to a decrease in the signal-to-noise ratio in the Doppler estimator.

Vice versa, we stress that all such signals should be subtracted in order to see the tSZ signal or other physical $y$-distortions in a clean way.  We have shown that the DD signal, which spans over all angular scales, is at most between $7\%$ and $15\%$ of the tSZ signal close to the direction of our dipole (see Figure~\ref{fig:DD-vs-tSZ}), and is less important in regions which are far away from it. This may not be a large contamination, but it is higher than the expected instrumental noise levels in the next-generation CMB experiments. For comparison the DQ is the largest distortion, but it only affects the $\ell=2$ mode.

Moreover such effects could contaminate measurements of intrinsic spectral distortions in the CMB:  while a monopole and a DQ are known to give rise to a $10^{-7}$ signal, we have pointed out that the DD gives rise to a non-negligible $10^{-8}$ signal on all multipoles. Even if one focuses on measuring only monopole distortions, also the latter should be carefully subtracted in order to avoid possible leakages due to partial sky coverage.

\section*{Acknowledgments}
We thank Barbara Comis for useful clarifications, and Craig Copi, Glenn Starkman, Jordi Miralda, Marcio O'Dwyer, Omar Roldan, Jim Zibin and Antony Lewis for interesting discussions. We also acknowledge important suggestions and corrections from the anonymous referees. MQ is grateful to Brazilian research agencies CNPq and FAPERJ for support and to the University of Barcelona for hospitality. AN is supported by the grants EC FPA2010-20807-C02-02, AGAUR 2009-SGR-168.

\appendix

\begin{figure}[b]
    \centering
    \includegraphics[width=0.47\textwidth]{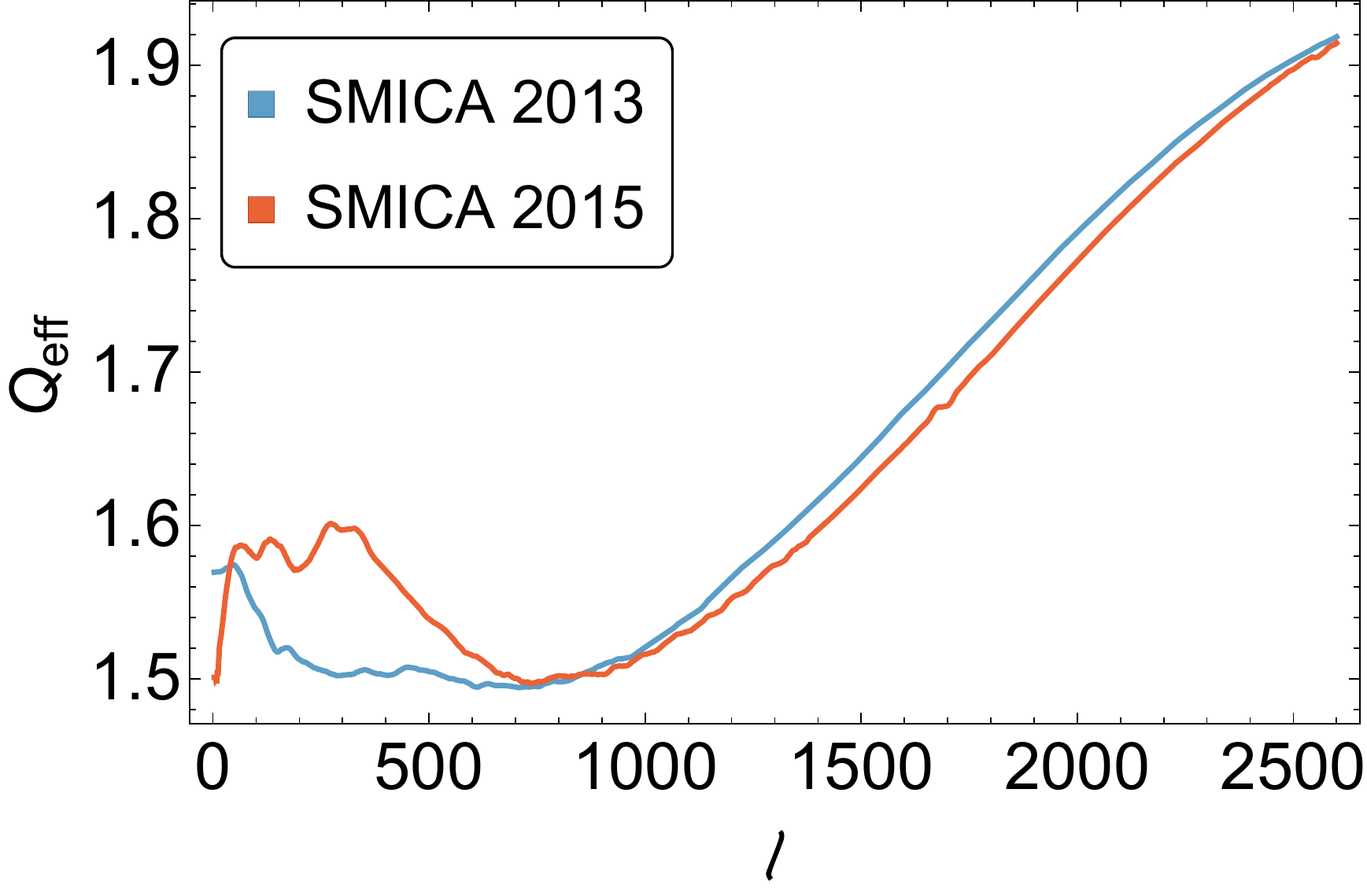}
    \caption{The weighted sum $Q_{\rm eff} \equiv \sum_i w_i Q(\nu_i)$ of the $y$-channel contribution to Planck SMICA 2013 and SMICA 2015 CMB temperature maps. See also Section~\ref{sec:detecting-DD}. \label{fig:smica2013-wQ}}
\end{figure}

\section{DD in Planck maps}\label{app:planck-Qw}

Planck maps were not built with the goal of removing the Dipolar Distortions.
In fact, their CMB temperature maps do not even project out the $y$-channel. This is probably due to the fact that the tSZ was not a large foreground and thus not removed in their different component separation techniques.  This can be explicitly seen for the SMICA maps. In particular, for SMICA 2013 and 2015 we computed explicitly the sum $Q_{\rm eff}=\sum_i w_i Q(\nu_i)$ of eq.~\eqref{eq:wQ=0} using the reported weights for all different multipoles. The result is in Figure~\ref{fig:smica2013-wQ}. The weights were reconstructed from~\cite{Ade:2013hta} and~\cite{Adam:2015tpy}. Note that the result is different from zero at all scales, meaning that there is a contamination due to the $y$-channel and thus also due to the DD. For the other map-making techniques used by Planck (NILC, SEVEM and Commander) defining an effective $Q$ for different scales is not a straightforward task as they are not obtained through a simple weighted sum in harmonic space.

\bibliographystyle{apsrev4-1}
\bibliography{cmb-new}

\end{document}